# Pixel-Level Calibration for Space-Based High Precision Astrometry Using Young's Fringes

Hugo Rousset *[a], Eric Thiébaut[c], Chloé Rey[a], Manon Lizzana[a,b,d], Sébastien Soler[a], Fabrice Pancher[a], Julien Michelot[b], Logi Olgeirsson[b], Jean Baptiste Le Bouquin[a], Alain Léger[a], Fabien Malbet[a]

[a]Institut de Planétologie et d'Astrophysique, 414 Rue de la Piscine, 38400 Saint-Martin-d'Hères; [b]PYXALIS, 170 Rue de Chatagnon, 38430 Moirans ; [c]Centre de Recherche Astrophysique de Lyon, 9 Avenue Charles André, 69561 Saint Genis Laval cedex, [d]Centre National d'Etudes Spatiales, 18 Avenue Edouard Belin, 31400 Toulouse

## ABSTRACT

The detection of Earth-like exoplanets via astrometry necessitates centroiding precision at the 0.3 µas level, thereby imposing stringent constraints on the focal plane geometry of space telescopes. The AGATE focal plane instrument, a proposed component of the NASA Habitable Worlds Observatory (HWO), aims to achieve this objective through the calibration of the intra-pixel response function of CMOS detectors to an accuracy of 50 µpix (220 nm The present paper proposes a Young's fringes-based calibration method, derived from those employed by JPL and IPAG, for the purpose of mapping the pixel response barycenter offsets $(\delta_x, \delta_y)$ across the detector. Using a Pyxalis GIGAPYX-4600 CMOS sensor, we demonstrate via simulations and laboratory measurements that (1) a precision of $5\times10^{-4}$ pix is achievable with 10,000 frames (current setup); (2) inter-pixel capacitance crosstalk and fringe hyperbolicity are dominating the errors at small and large scales, respectively; and, (3) an iterative inverse problem approach with a hyperbolic fringe model is proposed to overcome paraxial approximation limits for 1 Gpix focal planes. These results pave the way for on-board calibration of HWO's astrometric instrument, ensuring the sub-µas precision required for exo-Earth detection.

**Keywords:** astrometry, exoplanets, detector, pixel calibration, HWO, characterization, intra-pixel-response

## 1. INTRODUCTION

Relative astrometry has been demonstrated to be a highly effective tool for the detection of exoplanets, as it enables direct measurement of the reflex motion of a star resulting from the gravitational influence of an orbiting companion. In contrast to the limitations of radial velocity (RV) methods, which are subject to sin(*i*) degeneracy, where *i* denotes the orbital inclination, astrometry offers a distinct advantage in its capacity to provide unambiguous mass estimates for companions. For Earth-like planets at 10 parsecs, the astrometric signal is on the order of 0.3 µas, requiring centroiding precision at the $5\times10^{-5}$ pixel level (ratio of PSF radius and expected signal). This precision represents an improvement of approximately two orders of magnitude over the current capabilities of the Gaia telescope (10 µas), necessitating the development of novel calibration techniques for upcoming space telescopes such as the HWO.

In order to meet this requirement, it is necessary to calibrate not only the classical detector parameters (flat field, dark current, bias) but also the geometric misalignment of pixel response centroids. Manufacturing imperfections, such as photolithography and stitching, as well as intra-pixel response asymmetries (due to silicon defects or doping variations), introduce sub-pixel distortions that may introduce bias in centroid measurements. The metrology method employed utilizes Young's interference fringes as a ruler, enabling the mapping of distortions with picometric precision.

In this context, we propose two missions to detect Earth-like exoplanets orbiting Sun-like stars up to 10 parsecs (see Fig. 1). To achieve the desired precision, it is imperative to determine the star centroid with an accuracy of 0.3 µas[1]. This stringent requirement ($5\times10^{-5}$ pix for Habitable World Observatory astrometry instrument) necessitates the observation of the same star on multiple occasions and an in-depth understanding of the instrument. First, the optical distortion of the telescope must be accurately estimated. Subsequently, the detector parameters must be calibrated with precision to facilitate the extraction of star centroids. The primary and classical parameters include the response map, the dark current map, and the bias map.

*hugo.rousset1@univ-grenoble-alpes.fr

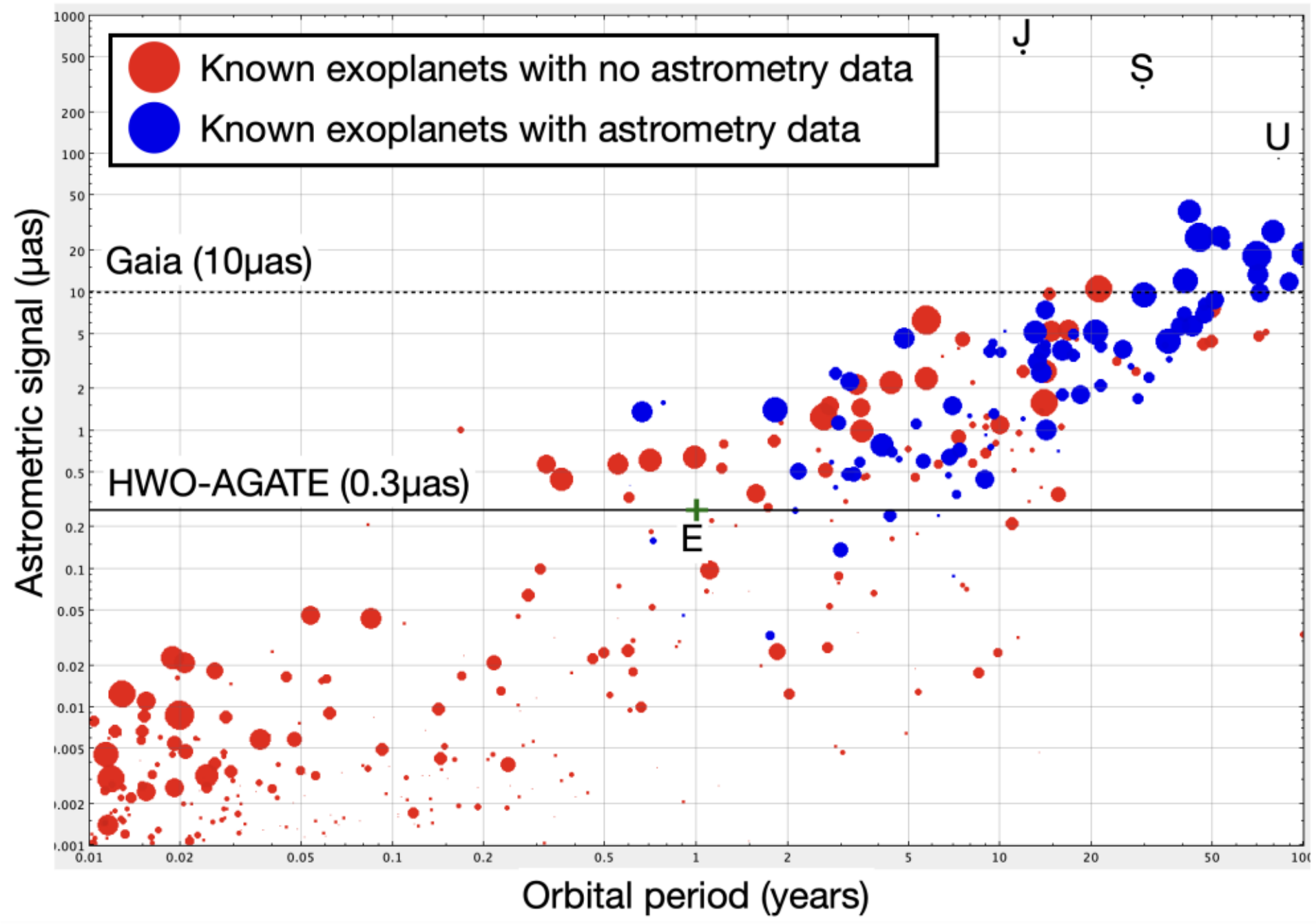

Figure 1. Astrometric signal from the known exoplanets registered in the Extrasolar Planets Encyclopaedia as a function of their orbital period. The size of the symbols is proportional to the mass of the exoplanets. Blue and red colors represent exoplanets with and without astrometric measurements, respectively. The two lines correspond to the Gaia detection limit (dashed line) at 10 µas and Theia/HWO-AGATE (solid line) at 0.3 µas. None of the exoplanets found before the Gaia DR4 2026 release can be detected with the Gaia detection limit, and only a few giant exoplanets can be detected with high astrometric precision.

However, to achieve the requisite precision, it is also necessary to calibrate the positions of the barycenters of the pixel responses. It is generally accepted that the pixel grid is perfectly spaced by the pixel pitch. However, when the manufacturer fabricates the matrix detector, several factors render this assumption inaccurate. Firstly, the geometric alignment of the pixels on the grid is imperfect due to photolithography and stitching inaccuracies. Secondly, the intra-pixel response may not be perfectly centered within the pixel due to defects in the silicon detection layer or variations in doping (Fig. 2).

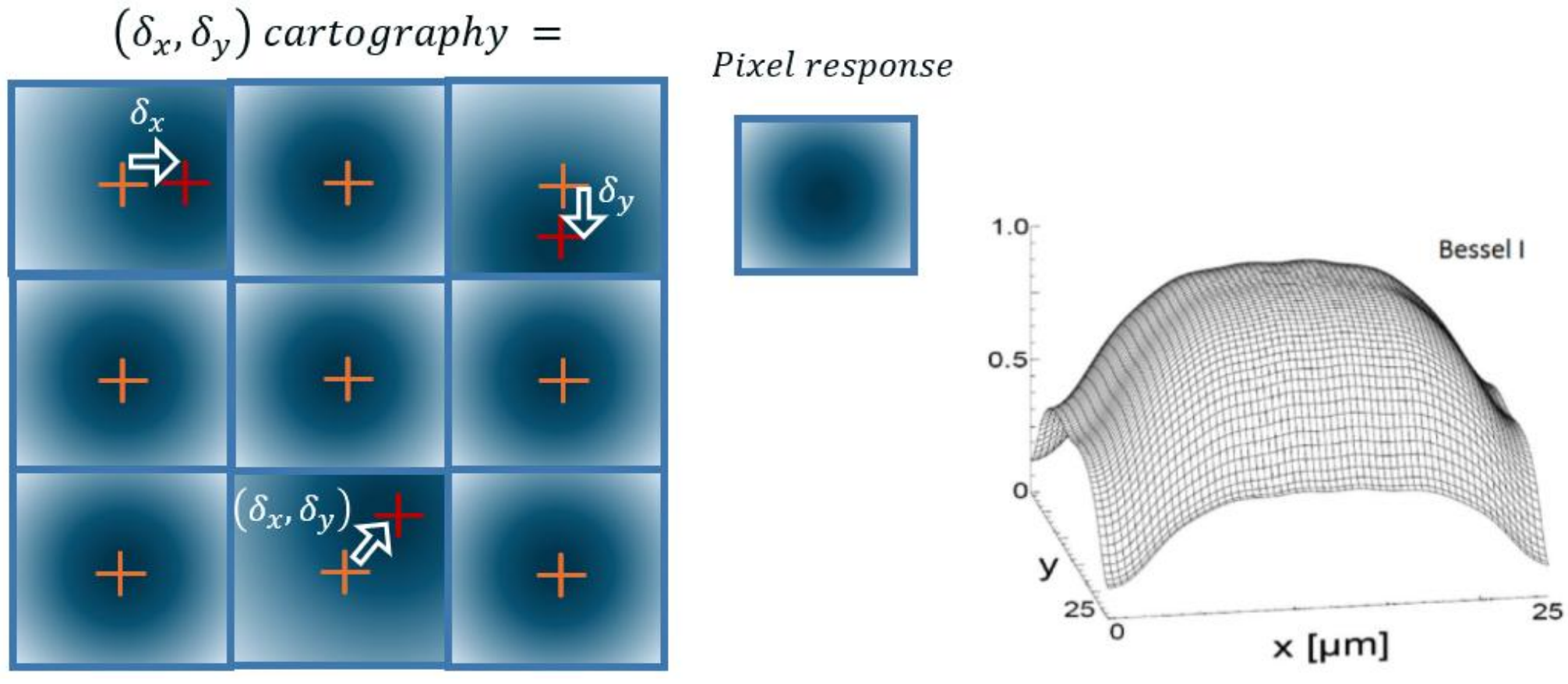

Figure 2. Left: Schematic representation of the $(\delta_x, \delta_y)$ parameters. Those represent the difference between a perfect grid at pixel pitch and the true response barycenter positions of each pixel. Right: Example of intra-pixel response of a CCD detector from Vorobiev (2019)[3]

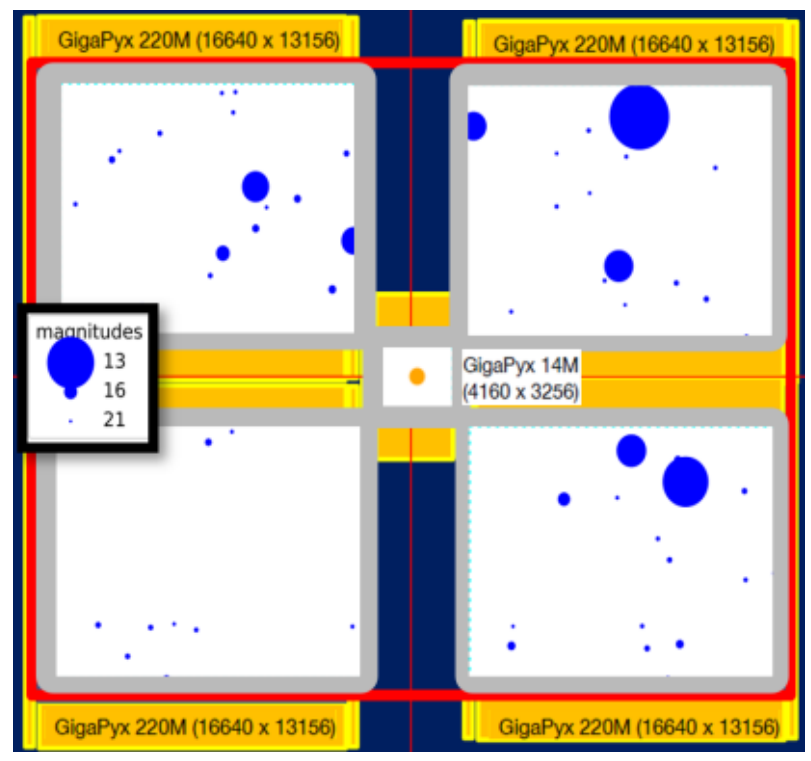

Figure 3. Focal plane contemplated for HWO-AGATE astrometry.

The present paper offers our inaugural findings on the calibration of the positions of the pixel response barycenters. The term $(\delta_x, \delta_y)$ is employed to denote the discrepancy between the true response barycenters and the ideal grid. The following expressions facilitate the representation of these concepts:[2]:

$$\delta_x = \frac{\iint xIPRF_\lambda(x,y)dxdy}{\iint IPRF_\lambda(x,y)dxdy} - x_0, \quad \text{and } \delta_y = \frac{\iint yIPRF_\lambda(x,y)dxdy}{\iint IPRF_\lambda(x,y)dxdy} - y_0 \qquad (1)$$

Where *IPRF* is the Intra-Pixel Response Function, depending on the light wavelength $\lambda$ and the integration is made on the whole pixel surface. A schematic view of those parameters is given in Fig.2. We will always express $(\delta_x, \delta_y)$ in fractions of pixels.

## 2. FOCAL PLANE ARRAY

The HWO (Habitable World Observatory) telescope has an entrance pupil of 7 m, a focal length of 130 m, thus a diffraction limit of 17 mas @ 600 nm, and a field of view of 4×3 arcmin². In order to properly sample the diffraction point spread function at Shannon-Nyquist, it is necessary that the pixel pitch of the detector be smaller than twice the PSF in the focal plane (Point Spread Function). This results in a pitch of no greater than 0.5 fλ/D, which is equivalent to 5.6 µm at 600 nm. The field of view of a pixel is 7×7 mas.

A review of the extant literature reveals that the two CMOS detectors with the proper pitch are the GIGAPYX from PYXALIS[3] and the Sony IMX411. Nonetheless, the IMX411 is equipped with a micro-lens array situated in front of the pixels, a configuration that is incompatible with the requirements of our particular application. The focal plane necessitates a minimum of 20×27 kpix² to adequately cover this field of view. The conceptual astrometry focal plane for HWO-AGATE is illustrated in Fig. 3. The primary objective of the 14-megapixel central detector is to acquire precise images of the science target, whereas the four 220-megapixel detectors are designed to facilitate the imaging of reference stars of lower brightness.

Table 1. Main performances of the GIGAPYX-4600

| Property | PYXALIS measurement |
|---|---|
| # Pixels | 8320×5456 |
| Pixel pitch | 4.4µm |
| Read Out Noise | $16e^-$ (in High Gain) |
| Linearity error | 2% |
| Response heterogeneity | <2% |
| Full well | >50ke$^-$ (in High Gain) |
| Dark current | $24e^-$ @ 35°C |

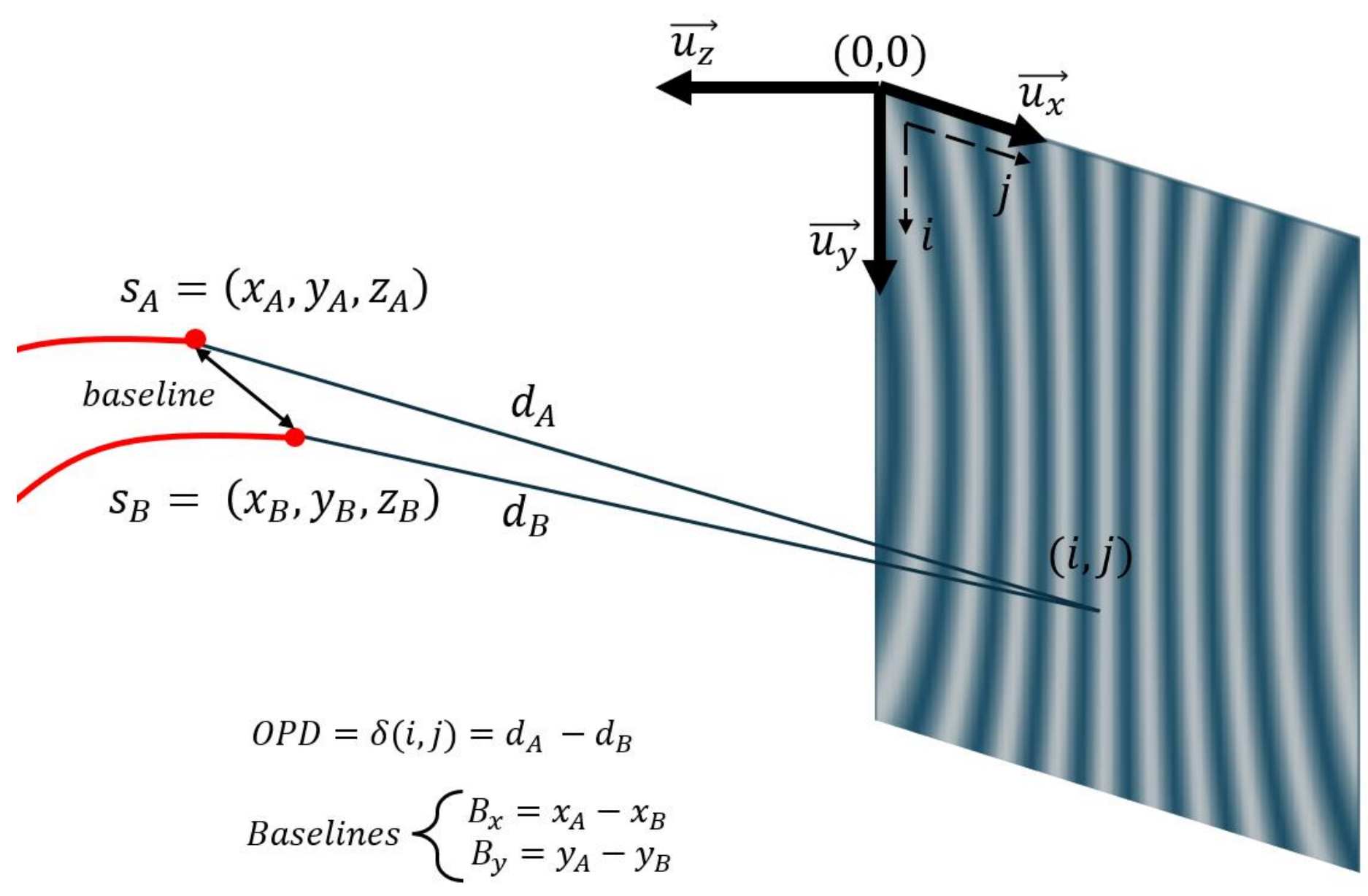

Figure 4. Geometrical conventions and notations for fibers and pixel positions.

The present study employed a PYXALIS GIGAPYX-4600[3], equipped with 46 megapixels. The primary performances are documented in Table 1. The detector is operated using the demo kit camera furnished by PYXALIS, known as GENEPYX. The same characterizations were performed at IPAG, and similar results were obtained[4].

According to the manufacturer's estimates, the precision of pixel alignment relative to a perfect grid is approximately $10^{-3}$ pixels (4.4 nm) between individual pixels and $4\times10^{-3}$ pixels (18 nm) between stitches.

## 3. MODELLING AND METHOD

The detector response barycenter map calibration that was implemented was derived from a NASA concept that was developed in 1995[5]. This on-board metrology concept was subsequently developed at JPL (Jet Propulsion Laboratory) by Mike Shao's team around 2012[6–9]. The concept was also proposed in a context of relative astrometry for exo-Earth detections. The objective is to execute the calibration process on-board with regularity to ensure the management of variations in the response barycenters maps during the mission.

### 3.1 Modelling

The fundamental principle underlying this process involves the placement of two optical fibers, each fed by coherent light, in front of the detector. Subsequently, interference fringes—also referred to as Young's fringes—will appear on the detector. The fundamental principle underlying this method is to utilize the equally spaced, well-known fringes as a ruler, thereby mapping their barycentric response. The most precise information will be located in the high gradient areas. Therefore, in order to adequately assess all the pixels, it is necessary to scan the fringes on the detector by temporally varying the dephasing between the two fibers. This is accomplished through the use of a phase modulator on one of the two fibers. It is imperative to note that at least two sets of fringes are required to achieve the necessary precision in the recovery of misalignments in both the horizontal and vertical directions of the detector. Ideally, these fringes should be orthogonal to each other (see Fig. 4).

Subsequent to the acquisition of the fringe data cubes, it is necessary to estimate the misalignments of the pixels $(\delta_x, \delta_y)$. To achieve this objective, the algorithm proposed by Antoine Crouzier in 2016[9] —which bears a notable similarity to the JPL[8] algorithm — was implemented. The objective is to estimate the parameters of a physical model that is representative of the fringes acquisition. The model is expressed as follows:

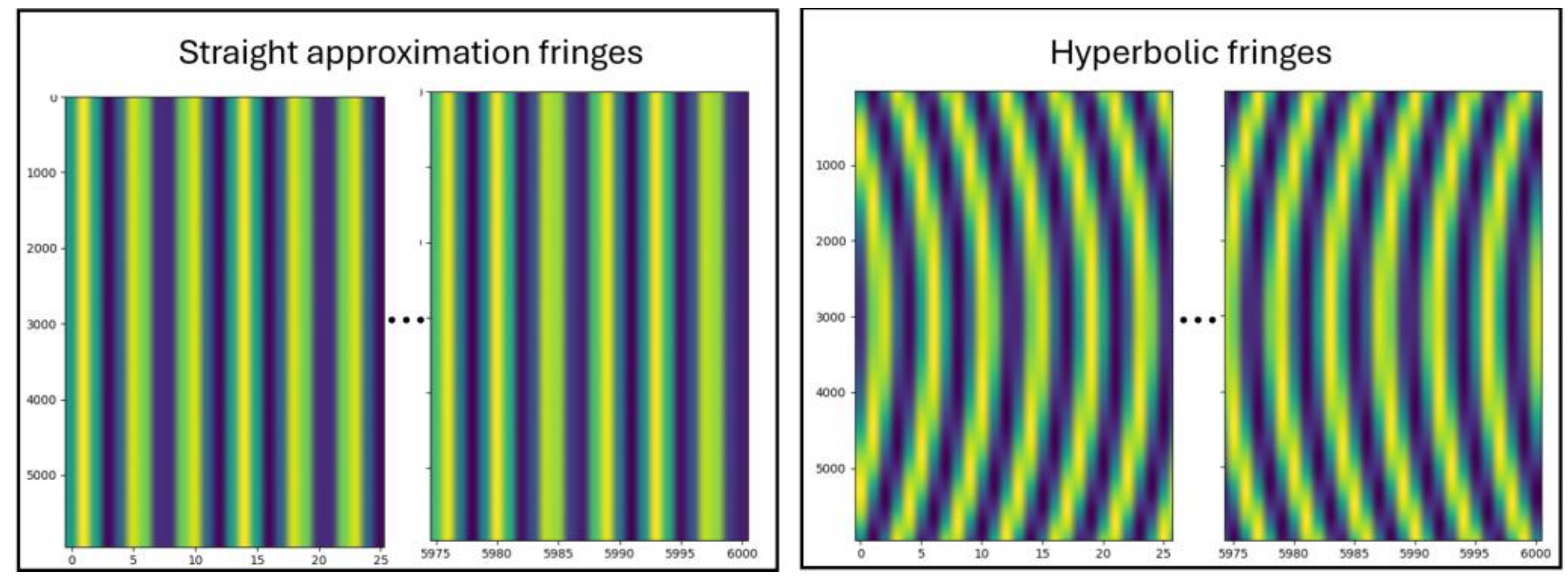

Figure 5. Straight or paraxial approximation versus real hyperbolic fringes simulation. This is for 40 cm distance fibers centered and 6000 × 6000 detector of 4.4 µm pixel pitch. Zoom on first 25 and last 25 columns.

$$I(i,j,t) = B(t)\iota(i,j) + A(t)\alpha(i,j) \times \sin\left(\frac{2\pi}{\lambda}\delta(i,j) + \phi(t)\right) \quad (2)$$

In this equation $I(i,j,t)$ represents the measured pixel values with regard to pixel number and acquisition instant, $\lambda$ is the illumination wavelength. Then there is a background term $B(t) \times \iota(i,j)$ representing the mean intensity light $I_A(i,j,t) + I_B(i,j,t)$ ($I$ being the two fibers intensities) multiplied by detector response, bias and dark residuals, straylight. $B(t)$ represents to first order the laser fluctuations and $\iota(i,j)$ (normalized) the mapping of the mean intensity. The second term $A(t)\alpha(i,j)\sin\big(\delta(i,j) + \phi(t)\big)$ represents the fringes themselves. $A(t) \times \alpha(i,j)$ is similar to the mean intensity parameter but for fringes contrast, thus including $2 \times \sqrt{I_A(i,j,t)\, I_B(i,j,t)}$, light polarization factors and light incoherence terms. The sinus represents the fringes shape with $\phi(t)$ the induced dephasing by the phase modulator of the bench and $\delta(i,j)$ the optical path difference between the two fibers to the pixel $(i,j)$. In this case, the fibers' positions are assumed to be constant over the course of time. Figure 4 exposes the conventions for the axis, fibers positions and pixel indexing.

The true expression of $\delta(i,j)$ is the simple difference of distance between fiber tip A and fiber tip B with regard to the pixel response barycenter. Thus:

$$\delta(i,j) = \sqrt{(x_A - (j+\delta_x)p\,)^2 + \big(y_A - (i+\delta_y)p\,\big)^2 + {z_A}^2} - \sqrt{(x_B - (j+\delta_x)p\,)^2 + \big(y_B - (i+\delta_y)p\,\big)^2 + {z_B}^2} \quad (3)$$

Where $p$ is the mean pixel pitch. This expression takes into account a probable tilt between the fiber baseline and the detector plane. $s_{A,B} = \big(x_{A,B},\ y_{A,B}, z_{A,B}\big)$ are the two fibers tips positions.

### 3.2 Paraxial approximation

We can show that in a paraxial approximation and considering $z_A = z_B = D$, the optical path difference can be written as follow:

$$\delta(i,j) = (j+\delta_x)K_x + \big(i+\delta_y\big)K_y \text{ with } K_x = \frac{pB_x}{D},\ K_y = \frac{pB_y}{D} \quad (4)$$

For the time being, this paraxial approximation is regarded as a reliable estimate, and its validity will be addressed subsequently. This approximation suggests that the fringes are perceived as straight on the detector, which is in contrast to the actual expression that yields hyperbolic fringes (see Fig. 5). This approximation is valid for detectors of small size, for which the pixels are sufficiently proximate to the projected fiber baseline center.

### 3.3 Method

The iterative method described by Crouzier et al.[9], was then implemented, and it consists of an iterative estimation process with two sub-iterations (see Fig. 6). In the first sub-iteration, the temporal parameters of the model (*B, A, $\phi$, $K_x$, $K_y$*) will be estimated using the full frame at each instant. In the second sub-iteration, the spatial parameters ($\iota$, $\alpha$, including $(\delta_x, \delta_y)$ at second iteration) will be estimated at each pixel along the temporal axis. The implementation is executed in the paraxial approximation, as previously outlined.

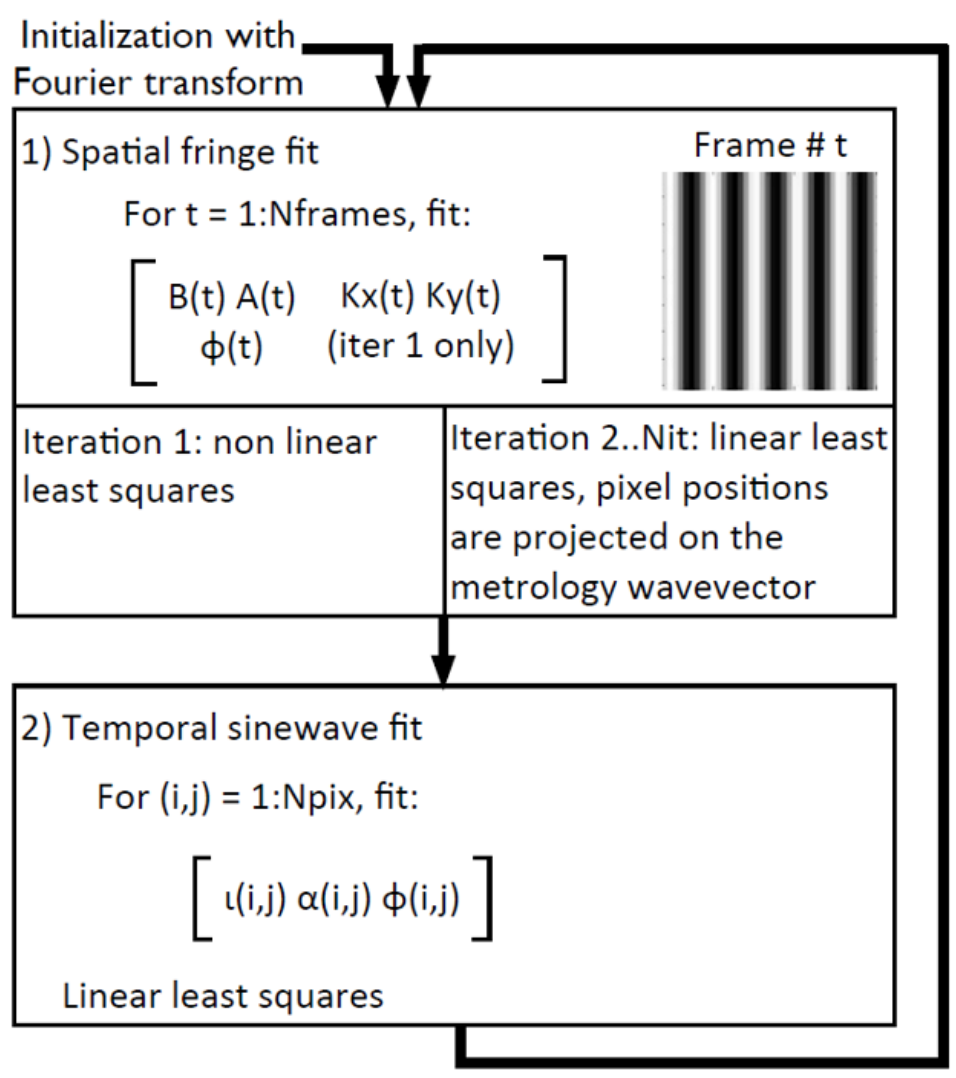

Figure 6. Iterative fringes parameters estimation scheme extracted from Crouzier et al.[9].

The parameters are estimated using an inverse problem approach based on the distance between the values of the pixels in the dataset and the values modeled. This results in the following non-linear least squares minimization:

$$\hat{\theta} = \text{argmin}_{\,\theta} \sum_{i \in \mathbb{I}} w_i (d_i - I(r_i, \theta))^2 \quad (5)$$

Where $d_i$ is the $i$-th pixel value in the observed image, $w_i$ the weights, $r_i$ the pixels positions and $I$ the model. $\theta$ represents the model parameters to estimate $\theta = \{B, A, \phi, K_x, K_y, \iota, \alpha, \delta_x, \delta_y\}$ in the case of the paraxial approximation implemented here.

### 3.4 State of the art

To the best of our knowledge, two similar Young's interference pixel positions calibration benches have been implemented: one at JPL (Jet Propulsion Laboratory) and one at IPAG (Institut de Planétologie et d'Astrophysique de Grenoble), from which part of our hardware is inherited. The two benches calibrate an E2V CCD39 detector with 80×80 pixels, a pixel pitch of 24 µm (5 times larger than the GIGAPYX-4600) and 200 $ke^-$ full well. Mike Shao's team at JPL has announced a $1\times10^{-4}$ pix precision[7] on the $(\delta_x, \delta_y)$ maps of the E2V CCD39, while Antoine Crouzier at IPAG announced $4\times10^{-4}$ pix precision. The latter employed Allan deviation and a comparison of two estimations with disparate bench configurations (different baselines) to attain this precision level. The two groups employ the paraxial approximation in their work.

The estimation of the gain yielded by the utilization of the aforementioned maps was conducted with the objective of refining the estimation of star centroids positions. This estimation represents the ultimate objective of the process.

## 4. IPAG IMPLEMENTATION AND MEASUREMENTS STABILITY

### 4.1 IPAG bench

An image of the bench is provided in Fig. 7 and the equivalent scheme is depicted in Fig. 8[10]. The experiment utilizes a HeNe 632 nm laser, which is employed to power a monomode fiber that employs phase-maintenance technology. The coherence length of the laser is 30 cm. The light is then separated into two fibers. Each fiber is directed to an electro-optic phase modulator, which is controlled by a function generator. This modulator is used to adjust the phase's amplitude, as desired. The configuration of the system is such that two outputs are connected to a 2×8 switch, which enables the redirection of each of the two inputs into four outputs. The switch is able to deactivate each output channel.

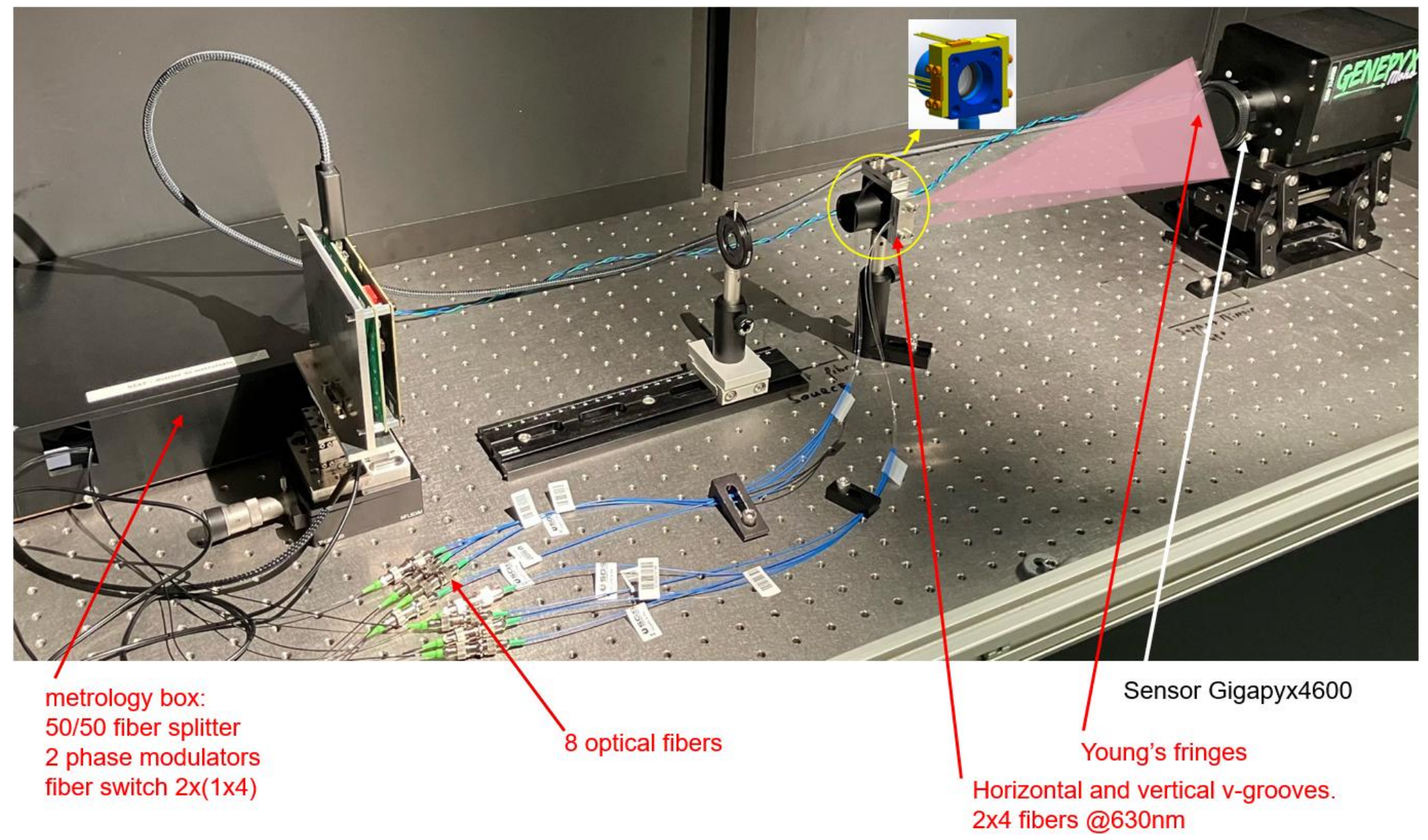

Figure 7. Interferometric calibration bench photographic.

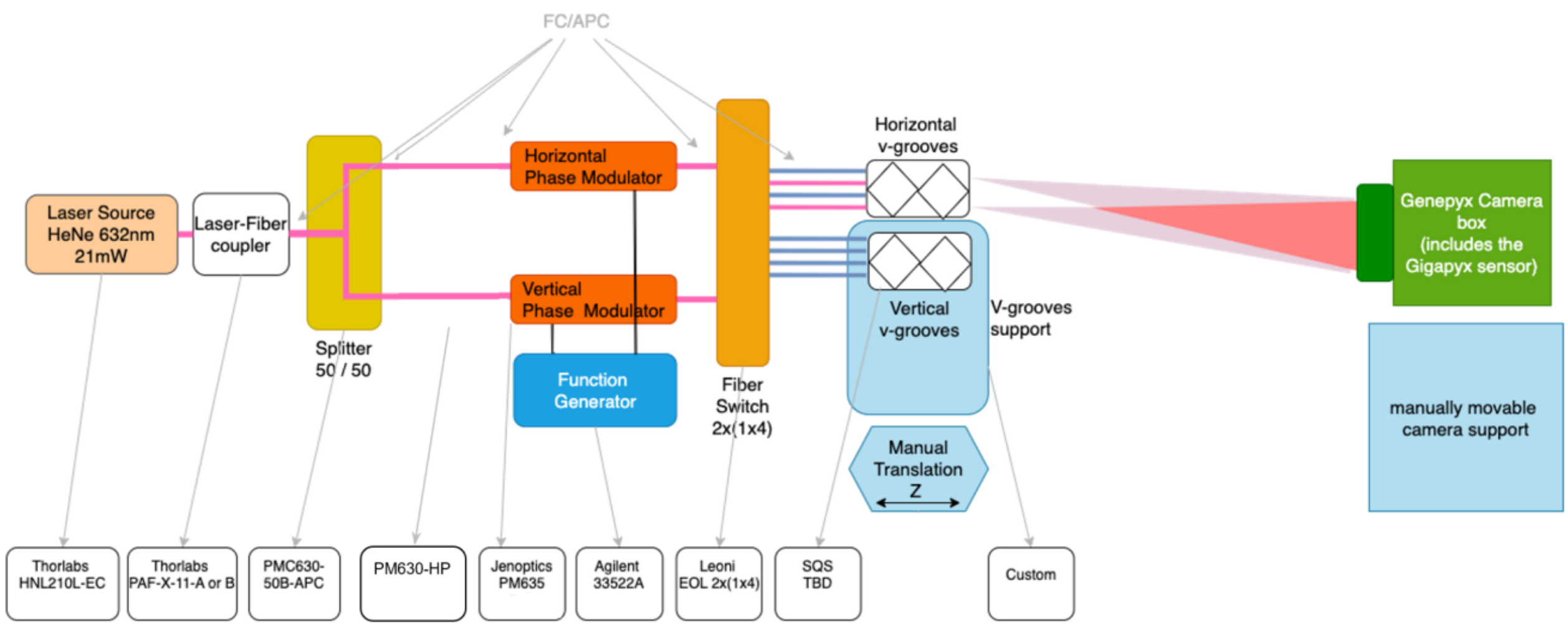

Figure 8. Scheme of the current interferometric bench

Subsequently, the eight optical fibers are arranged in two V-grooves, with each V-groove containing four fibers. One V-groove is positioned in a vertical orientation, with the objective of generating horizontal fringes, while the other is placed in a horizontal position, with the intention of yielding vertical fringes. Two fibers per V-groove would be adequate; however, four fibers would allow for the selection of six baselines for experimental purposes. The baselines exhibit unequal spacing, ranging from 1 mm to 6 mm. This variation corresponds to a range of 7 pixels to 57 pixels of interference at a distance of 40 cm. Following the initial measurement with the protective glass of the camera, the entire bench was transferred to a clean room environment. This was done to remove the protective glass, thereby ensuring that there was no intervening material between the fiber's tips and the pixels of the detector. The glass induced detrimental interference artifacts that proved impossible to mitigate through cleaning.

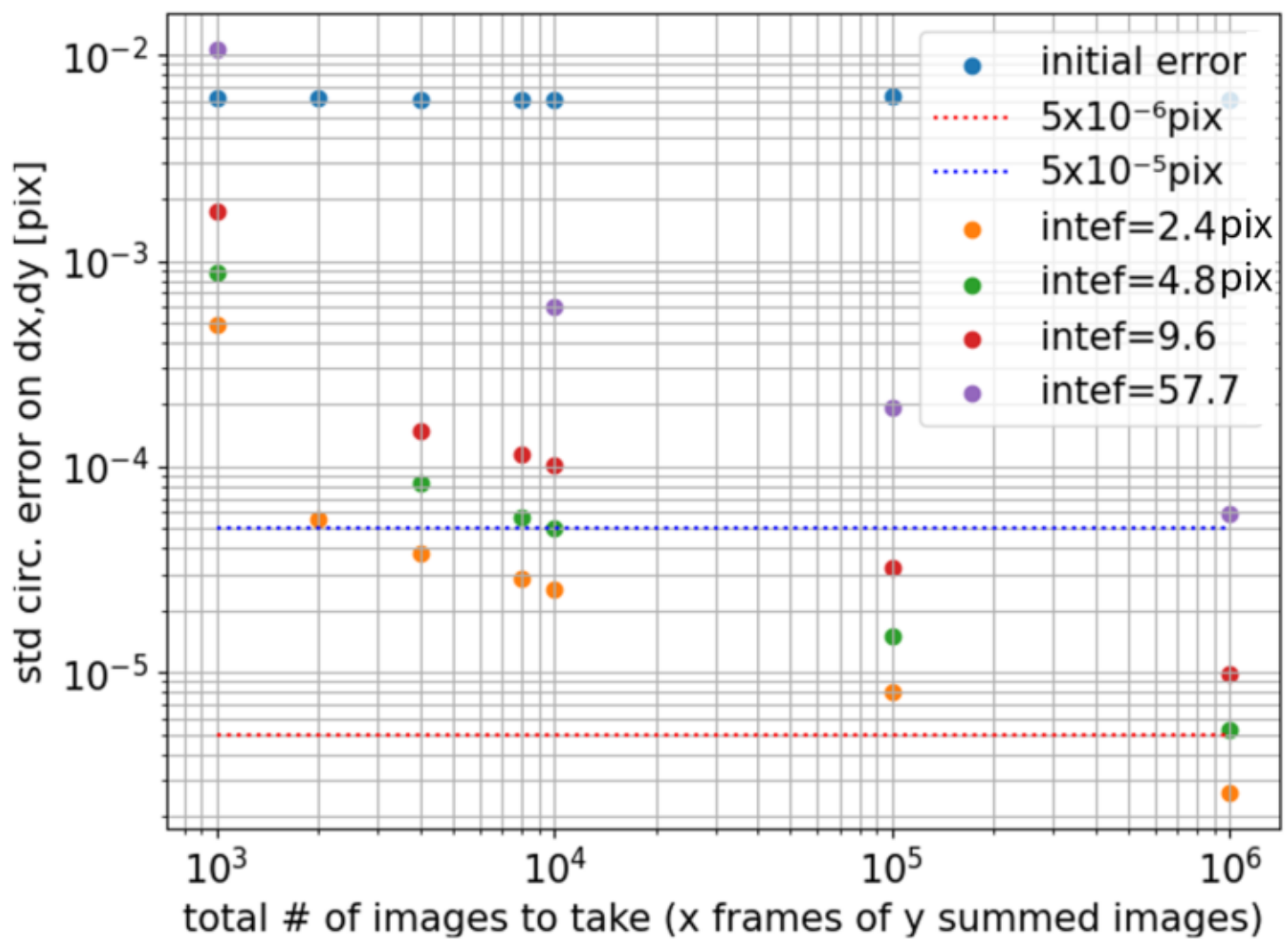

Figure 9. Simulation results of $(\delta_x, \delta_y)$ estimation precision versus number of integrated frames of 50 ke$^-$ The different colors correspond to different interfringes, thus different baseline/distance ratios.

### 4.2 Simulation of the fringes acquisition and parameters estimation

Initially, simulations were conducted to assess the validity of the implementation and the number of frames necessary to attain the desired precision for the mission ($5\times10^{-5}$ pix = 220 pm).

The simulation is characterized by its simplicity, with constant baseline and distance in the course of time, constant illumination flux, and homogenous response and dark current maps. The $(\delta_x, \delta_y)$ are randomly chosen in a gaussian distribution of deviation $6\times10^{-3}$ pix. The fibers illumination map is considered to be gaussian. The fibers illumination map is considered to be gaussian. The shot noise is incorporated into the simulated data as $\sqrt{e^{-}\text{count}}$ and read out noise of 15 e$^-$. $\phi(t)$ is randomly chosen at each frame.

As illustrated in Fig. 9, the estimation precision obtained on the $(\delta_x, \delta_y)$ is contingent upon the number of frames taken and the size of the interfringe (spatial period of the fringes). In order to achieve a reduction in computation time, the number of frames was decreased while simulating the accumulation of frames. This was accomplished by reducing the shot noise above $10^4$ frames. This would be analogous to summing frames of the same dephasing in real life.

As anticipated, the increased number of frames captured and the reduced interfringe result in enhanced precision. In theory, the smallest interval is at the Shannon-Nyquist limit, thus 2 pix. To ensure a margin of safety, the 5-pixel interframe case is considered for the on-board processing sizing at this time.

It is important to note that we attempted two distinct approaches in our estimation process: first, we allowed the temporal parameters $(A, B)$ to vary with time; second, we forced them to a mean constant value. The convergence and the precisions presented in Fig. 9 have been obtained by imposing a constant value on the parameters under consideration. When these parameters are released, the convergence becomes uncertain, and the necessary precisions are not attained. The metrology parameters $(K_x, K_y)$ are constant over time.

The process was also simulated with the utilization of measured cartographies of detector biases and responses that were acquired during the dedicated characterization. The precision values obtained in this case are consistent with those reported before. In this study, the results are limited to the case of vertical fringes, and the $\delta_x$ parameters are presented. It is noteworthy that the behavior exhibited by horizontal fringes is analogous. Figure 10 (left) shows the model estimated parameters after 5 iterations for this simulation. The $(A, B, K_x, K_y)$ parameters are constrained to constant values, while $\phi(t)$ is random which is in accordance with our simulation parameters. The bias and response maps appear in the $(\alpha, \iota)$ maps and the $\delta_x$ map is found to be spatially random as simulated. As illustrated in Fig. 10 (right), the estimated parameters are not subject to any systematic errors.

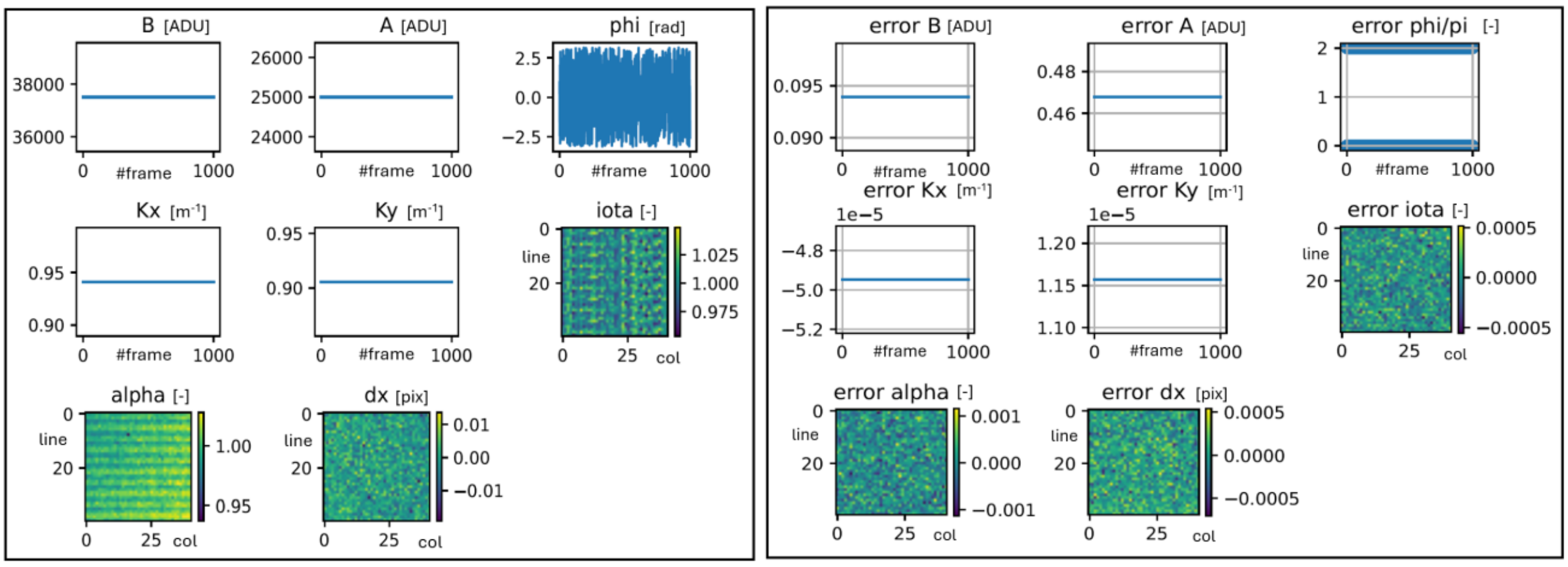

Figure 10. Left: model parameters estimated for simulation of 1000 frames with shot and read out noises, real bias and response cartographies. Temporal parameters are forced constant at beginning of the second iteration except for $\Phi(t)$. Right: errors on the estimated parameters with regard to the simulated ones. Simulation is conducted on vertical fringes to obtain the horizontal misalignments $\delta_x$.

Subsequent research will center on the algorithm's ability to maintain its efficacy as an estimator when faced with varying temporal $(A, B, K_x, K_y)$ parameters. Indeed, $(A, B)$ will be subject to laser intensity variations, which are estimated to 2 % for our hardware. Furthermore, the parameters $(K_x, K_y)$ are susceptible to thermal deformations of the bench, low-frequency vibrations, and other factors. Subsequently, it will be imperative to evaluate the repercussions of micro-vibrations that arise during the integration of a frame. These vibrations have the potential to induce image blur and fringes distortion.

### 4.3 Real measurement stability

The detector is acquiring at approximately 7 Hz (140 ms period) due to unoptimized software of the demo-kit camera. The integration time is approximately 20 ms for a 50 ke$^-$ full well. The detection of a sufficient number of photons requires a considerable amount of time. Therefore, we conducted a series of dedicated measurements, acquiring 50 kframes, to perform Allan deviation studies[9] and detect any long-term instabilities that might degrade the parameters estimation. The acquisition of 50,000 frames at a rate of 7 Hz requires approximately two hours.

Allan deviation batches contain 500 frames each and the $(\delta_x, \delta_y)$ estimation is computed separately on each batch (one $(\delta_x, \delta_y)$ map per 500 frames). Subsequently, the Allan deviation between those $(\delta_x, \delta_y)$ estimations is calculated. Figure 11 shows the two Allan deviation plots for two orthogonal diagonal fringes datasets. We chose diagonal fringes in order to decorrelate residual fringes patterns and electronic effects in the estimated parameters maps.

We see that our bench seems stable to extract $(\delta_x, \delta_y)$ maps up to 30 min measurement. An increase in the Allan deviation at 30 min for the second dataset is observed. This could be attributed to a long-term drift of a physical component of the bench. This drift is not evident in the first dataset, which suggests that it may not be a standard behavior.

Prior 30 min (13 kframes at 7 Hz) the two datasets behave well with square root law within the error bars. Therefore, it can be assumed that with a reduced interfringe (10 pixels in this datasets) a precision of $5\times10^{-4}$ pixels could be attained (10,000 frames for 5 pix interfringe, based on the previously presented simulation).

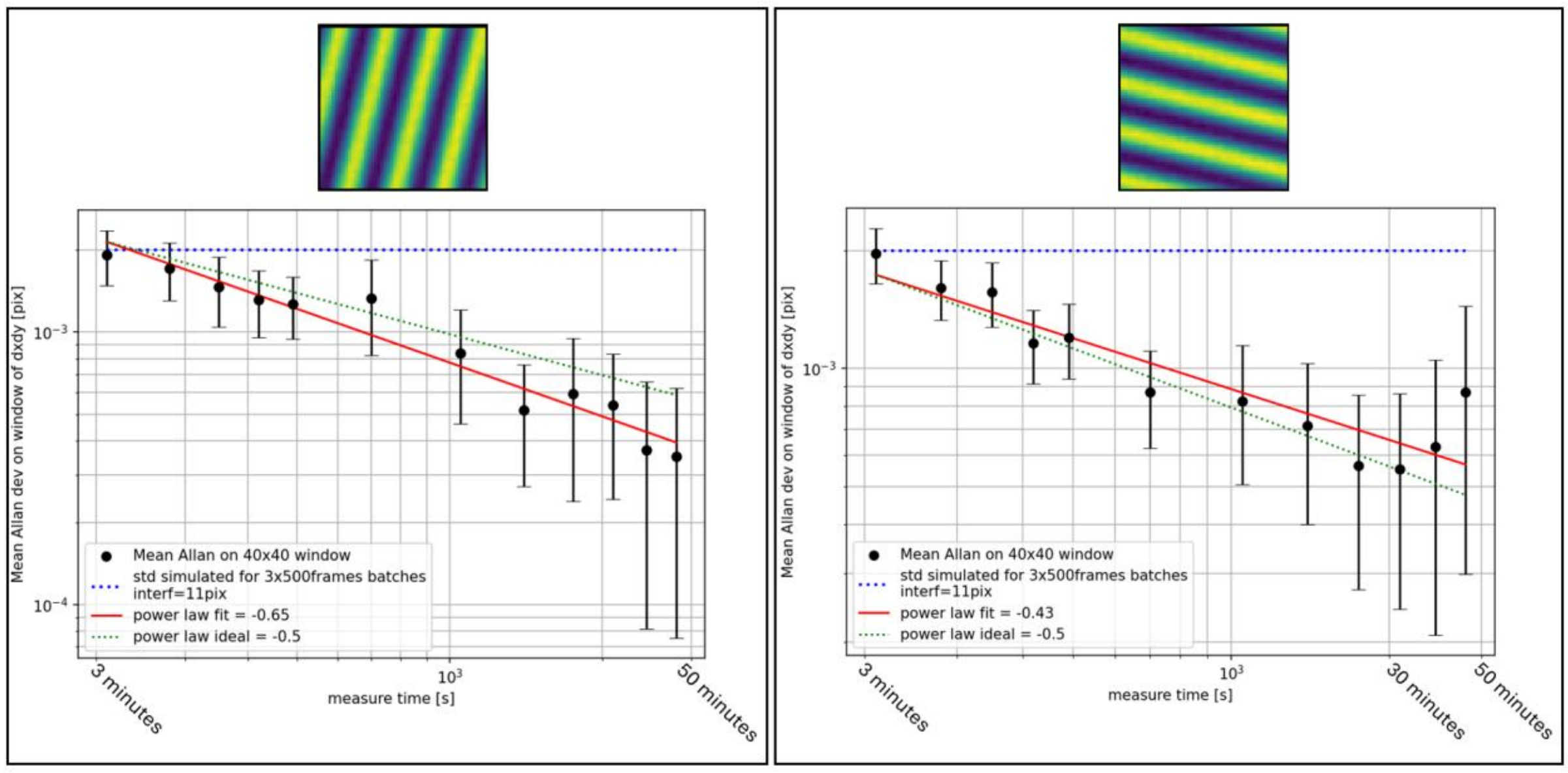

Figure 11. Allan deviations of $(\delta_x, \delta_y)$ (actually the norm on the fringe axis) for two datasets (fringes frame examples on the top 10 pix interfringe). Black dots are the average Allan deviations on the 40×40 window, red line is the power law fit, blue dotted line is the simulated standard deviation with shot noise for 1500 frames and similar interfringe than measurement, green dotted line is a perfect square root law.

## 5. FIRST ANALYSIS

A series of measurements were executed on the bench to estimate $(\delta_x, \delta_y)$ on specific regions of the GIGAPYX-4600. In this study, the term "small scale" is defined as a few tens of pixels, while "medium scale" is defined as a few hundreds of pixels.

### 5.1 Small scale reveals electronic crosstalk patterns

Two 10 kframe measurements were performed on the GIGAPYX-4600, one with vertical fringes and one with horizontal fringes. The $(\delta_x, \delta_y)$ were then estimated using our algorithm. The integration times are approximately 20 ms, and the fullwell of the detector is 50 ke$^-$. The results of this study are illustrated in Fig. 12. The obtained amplitudes of vertical misalignments are approximately 2 % of pixels, while those of horizontal misalignments are around 0.5 % of pixels.

It is observed that no fringes patterns persist on the $(\delta_x, \delta_y)$ maps. In the event that a reduced number of frames is utilized, the presence of fringes on the maps is typically observed. Following a series of simulations, we ascribe this phenomenon to the $(K_x, K_y)$ variabilities, representing geometric variations in the relative positions of the fibers and the detector, or alternatively, variations in the light wavelength. To substantiate our findings, it is imperative that we average these effects across a sufficient number of frames.

A vertical pattern characterized by saw-toothed shapes is observed in 4 % of pixels on the $\delta_y$ map and a similar pattern is seen with 0.2 % of pixels on the $\delta_x$ map. The amplitudes are too high to be attributed to geometrical misalignments of the pixels of the detector. Following a series of investigations, the observed effect was attributed to electrical crosstalk of the ROIC (read-out integrated circuit) of the GIGAPYX-4600. We performed dedicated measurements of the IPC[11] (Inter-Pixel Capacitance) effect of the detector using hot pixels[12,13] from dark current measurements.

IPC is defined as a capacitive effect that occurs between pixels or readout circuits of pixels. This effect implies that signal is exchanged between adjacent pixels. The IPC kernel measurements demonstrate a four-line repeatability, with a substantial contribution to the vertical axis (approximately 2-3 % of the impact of a pixel on the upper-lower one) and a less significant effect on the horizontal axis (around 0.6 % of a pixel on the left-right ones). The IPC[4] kernel is sparse, and the pixels impacted are not always the closest neighbors for the GIGAPYX-4600.

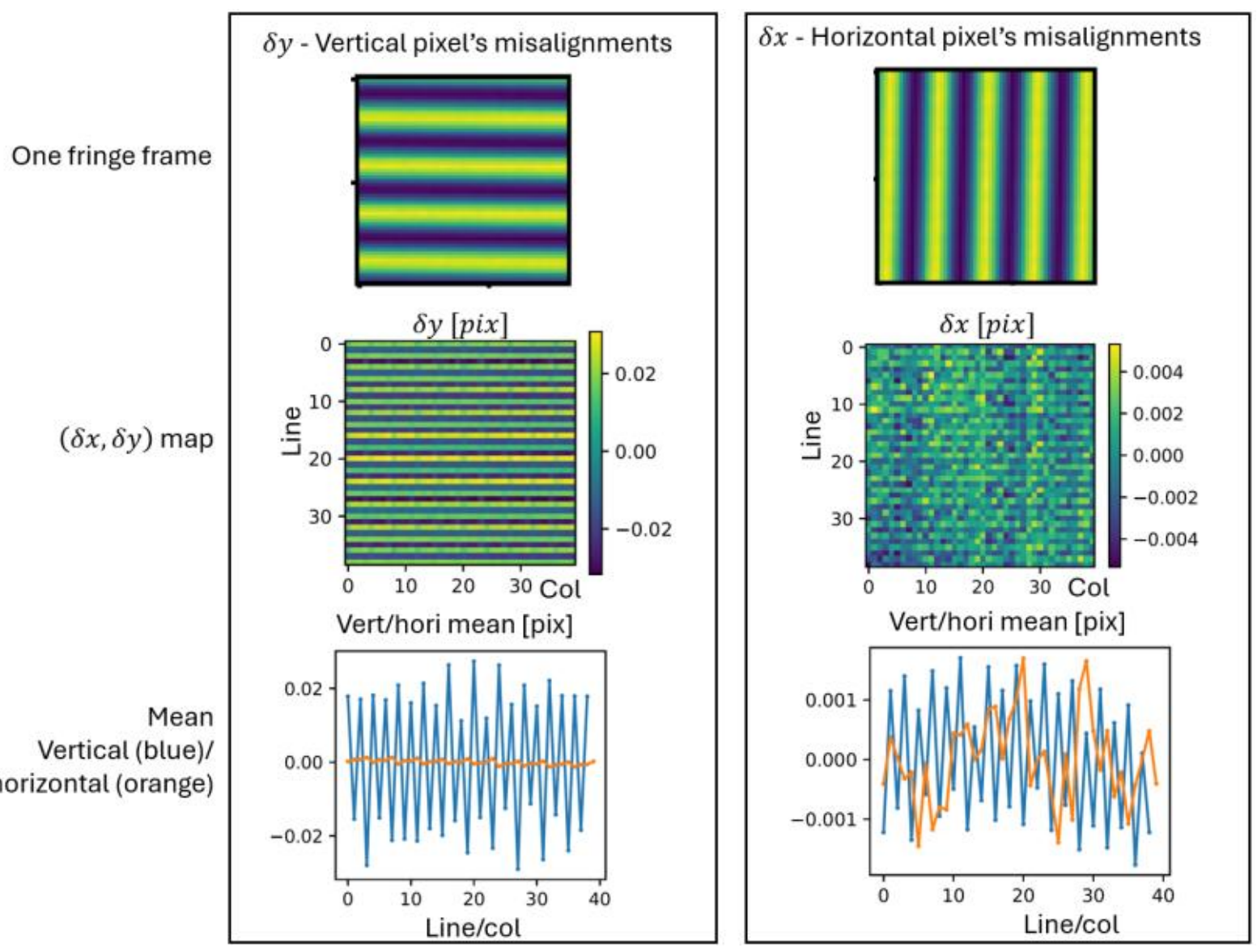

Figure 12. Real $(\delta_x, \delta_y)$ estimation for a 40×40 pix window on the GIGAPYX-4600 using 10 kframe – First line : one fringe frame in the 10 kframe – Second line : maps of $(\delta_x, \delta_y)$ – Third line : mean along lines and columns of the maps – Left: Vertical misalignments $(\delta_y)$ - Right: Horizontal misalignments $(\delta_x)$. Interfringe is 9 pix.

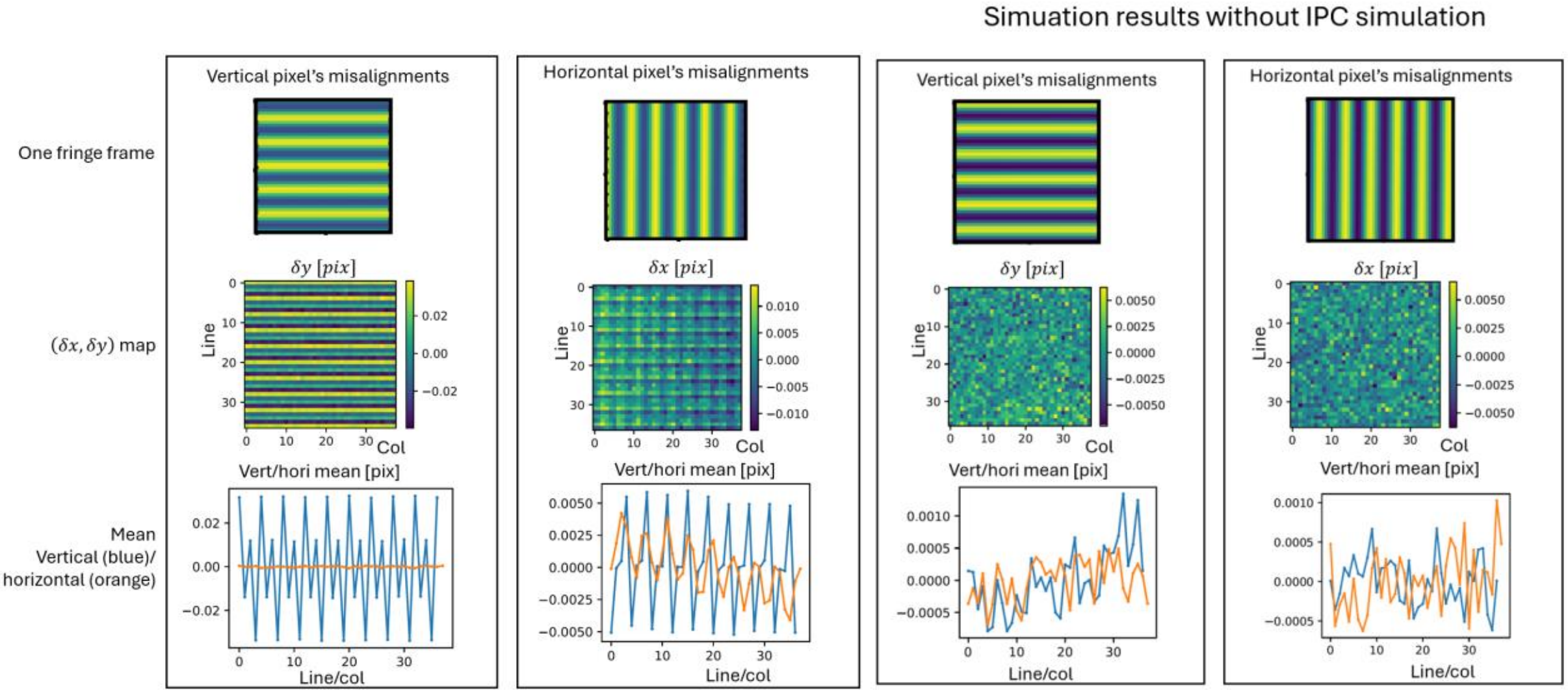

Figure 13. Simulation of $(\delta_x, \delta_y)$ estimation including measured IPC kernel impact for a 40×40 pix window on the GIGAPYX-4600 using simulated frame (no shot noise) – First line : one fringe frame – Second line : maps of $(\delta_x, \delta_y)$ – Third line : mean along lines and columns of the maps – Left: Vertical misalignments $(\delta_y)$ - Right: Horizontal misalignments $(\delta_x)$. Simulated $(\delta_x, \delta_y)$ are spatially distributed with $2\times10^{-3}$ pix standard deviation gaussian law, interfringe is 8 pix. Right panel: Same simulation without IPC.

Following the estimation of an IPC[4] kernel of the GIGAPYX-4600 employing the hot pixel method, simulations were conducted to elucidate its impact on $(\delta_x, \delta_y)$ maps. The results of these simulations are illustrated in Fig. 13. A comparison of the real misalignments and the simulated ones with IPC reveals a clear similarity between them. The saw teeth shape is reproduced, and the amplitudes are similar, particularly for the vertical misalignments. Horizontal ones exhibit higher amplitudes and slightly divergent shapes. The disparities observed between the actual and simulated data are attributed to an imperfect measurement of the IPC kernel. Indeed, our hot pixels measurement is noised and implies to change the detector chronograms with regard to fringes acquisitions, thus potentially different temporal crosstalks.

Figure 13 right panel shows the same simulation without the IPC emulation. The $(\delta_x, \delta_y)$ maps estimated are flat and the estimation precision goes down to $10^{-7}$ pix precision for $(\delta_x, \delta_y)$ maps, spatially distributed with $2\times10^{-3}$ pix standard deviation gaussian law. The simulations are executed with the averaging of the $(A, B)$ parameters. Algorithms to deconvolve IPC are under investigations.

### 5.2 Medium scale reveals fringes hyperbolicity patterns

At larger scales, the IPC's teeth shape remains consistent, and an additional pattern is superimposed. This phenomenon was referred to as the "paraxial approximation crest." Figure 14 shows the $(\delta_x, \delta_y)$ estimations of real measurements (sub-part of the dataset presented for the 40×40 pix windows). In the estimation of 40×40 pixel windows, 10,000 frames were utilized. However, due to constraints related to computational time, the medium scale study employs 500 frames. We can distinguish three phenomena:

- IPC: The vertical averages (blue) clearly demonstrate the impact of the IPC presented in the "small scale" paragraph, exhibiting the same crest-to-crest amplitudes.
- Fringes residuals: The maps demonstrate the presence of fringes residuals, characterized by horizontal 9-pixel period sinusoids on the $\delta_y$ map and vertical 9-pixel period sinusoids on the $\delta_x$ map. This phenomenon is attributed to the association of a low number of frames and time-varying geometry on the bench, i.e. $(K_x, K_y)$ variations due to vibration, thermomechanical deformations, etc.
- Paraxial approximation crest: The maps illustrate a diagonal crest, highlighted by the mean profiles. The presence of the phenomenon under consideration is primarily observed on $\delta_x$, although it can also be observed on $\delta_y$ depending on the fibers' positions. If we estimate the $(\delta_x, \delta_y)$ on a nearby windows including half of the current one and another half, the pattern remains the same, meaning that this is not a true signal but a post-processing artifact due to an imperfect intensity model. The line of crest direction and the curvature of the mount are determined by the misalignment of the fibers with respect to the detector area under study. On the $\delta_y$ map, the effect is negligible, likely attributable to two factors. Firstly, the center of the baseline of the fibers is in close proximity to the studied pixels. Secondly, the higher amplitude of the IPC phenomenon masks the paraxial approximation effect.

In order to understand the paraxial approximation crest effect, simulations were conducted. The fringes frames were simulated without approximation, thus taking the hyperbolic shape of the fringes into account. Subsequently, we estimated the $(\delta_x, \delta_y)$ maps using the paraxial approximation model, also known as the straight fringes model. Figure 15 shows the simulation results and a rescaled accordance with real data (right). The term "fibers offset" is employed to denote the offset between the projected center of the baseline on the detector and the center of the pixels window under study. We employed a trial-and-error approach with the fibers offset simulated to approximate the impact of the phenomena on the real data maps, given that we have not yet obtained the true fibers' position. The fibers offset measurements obtained are consistent with the precision of our laboratory's equipment, exhibiting a deviation of less than 2 centimeters for misalignment and a separation of 2 centimeters between the vertical and horizontal baselines. In order to incorporate the hyperbolic shape of the fringes in the model, it is necessary to estimate the positions of the fibers $s_{A,B} = (x_{A,B}, y_{A,B}, z_{A,B})$ instead of the approximated $(K_x, K_y)$ parameters.

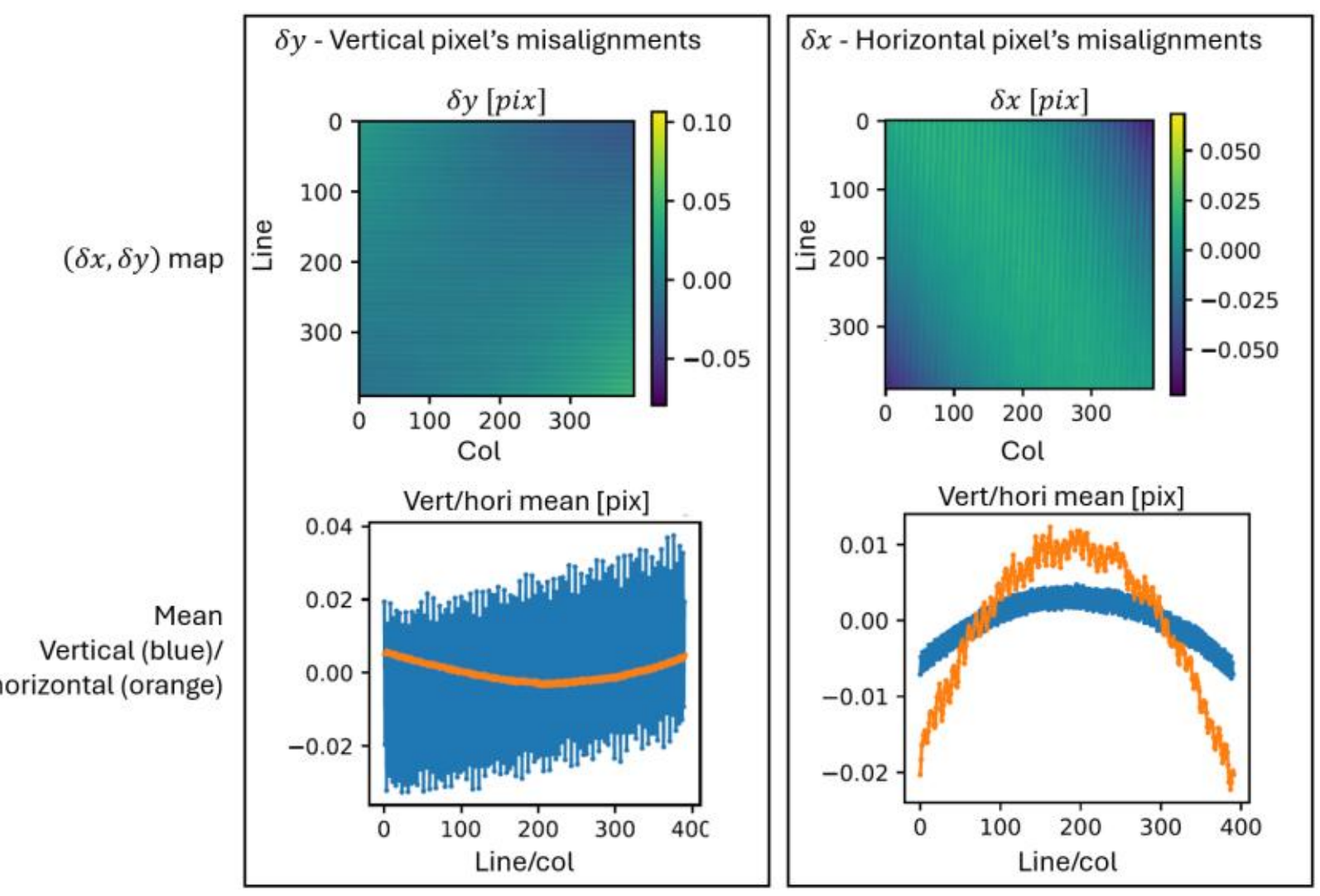

Figure 14. Real $(\delta_x, \delta_y)$ estimation for a 400×400 pix window on the GIGAPYX-4600 using 500 frames –First line: maps of $(\delta_x, \delta_y)$ – Second line: mean along lines and columns of the maps – Left: Vertical misalignments $(\delta_y)$ - Right: Horizontal misalignments $(\delta_x)$. Interfringe is 9 pix.

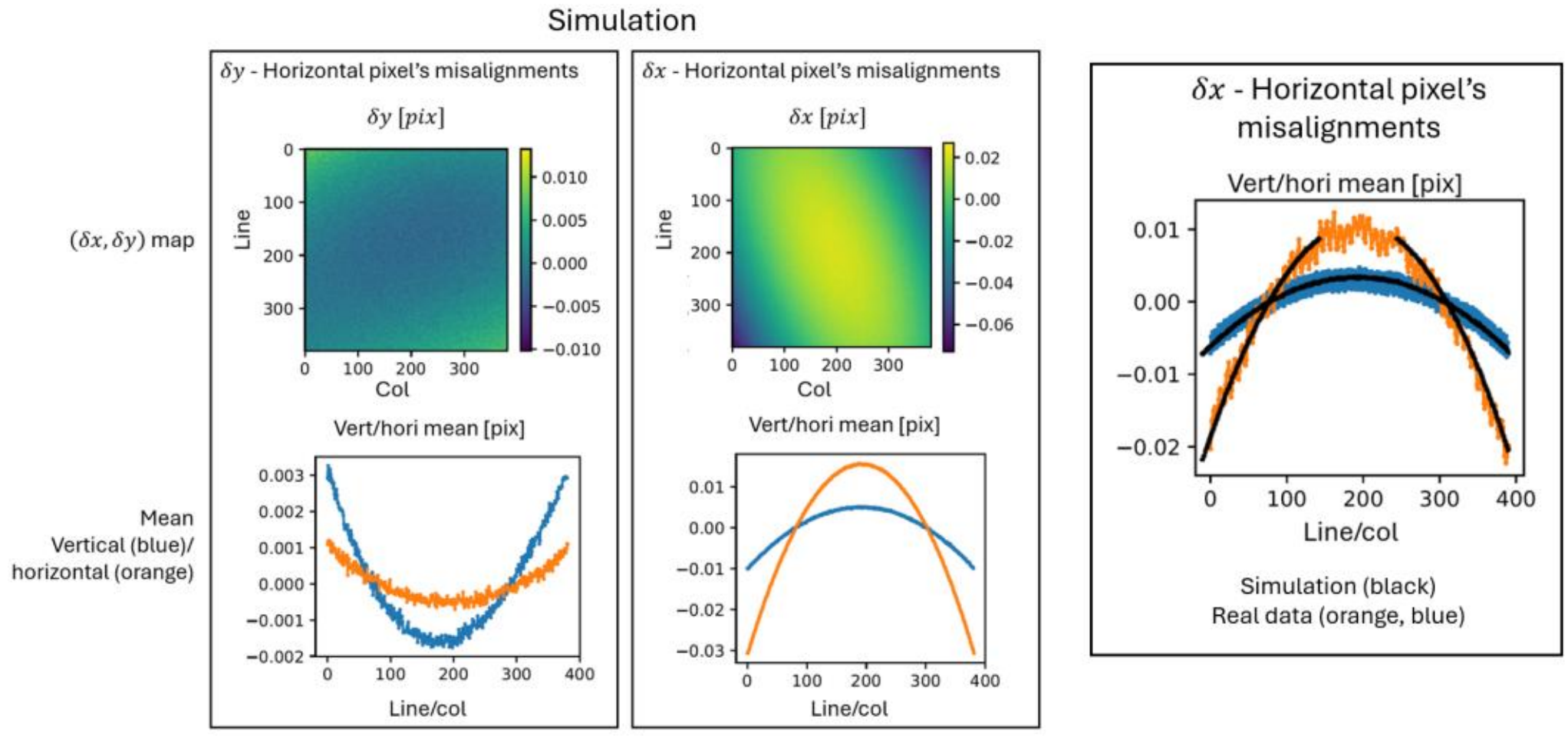

Figure 15. Simulation $(\delta_x, \delta_y)$ estimation for a 400×400 pix window on the GIGAPYX-4600 (no shot noise). Fringes simulated with full hyperbolic model, estimation with straight fringes approximation – First line: maps of $(\delta_x, \delta_y)$ – Second line: mean along lines and columns of the maps – Left: Vertical misalignments $(\delta_y)$, fibers offset (18 mm, 18 mm) - Right: Horizontal misalignments $(\delta_x)$, fibers offset (1.8 cm, -1.8 cm). Farther right: $\delta_x$ averages simulated (black, rescaled) and real data ones (blue, orange).

## 6. HYPERBOLIC FRINGES SOLVING

In an effort to directly implement the $(\delta_x, \delta_y)$ iterative estimation by inverse approach with the true hyperbolic model, we tried to estimate the fibers' positions $(x_{A,B}, y_{A,B}, z_{A,B})$ instead of the $(K_x, K_y)$ parameters in the first sub-iteration. This was done by solving the non-linear OPD (optical path difference) $\delta(i,j)$ formulation. However, the inverse problem approach currently employed is excessively computationally intensive and does not converge when initial estimates of fiber positions are set to a reasonable value. This phenomenon can be attributed to the presence of multiple deep secondary minima within the inverse problem cost function.

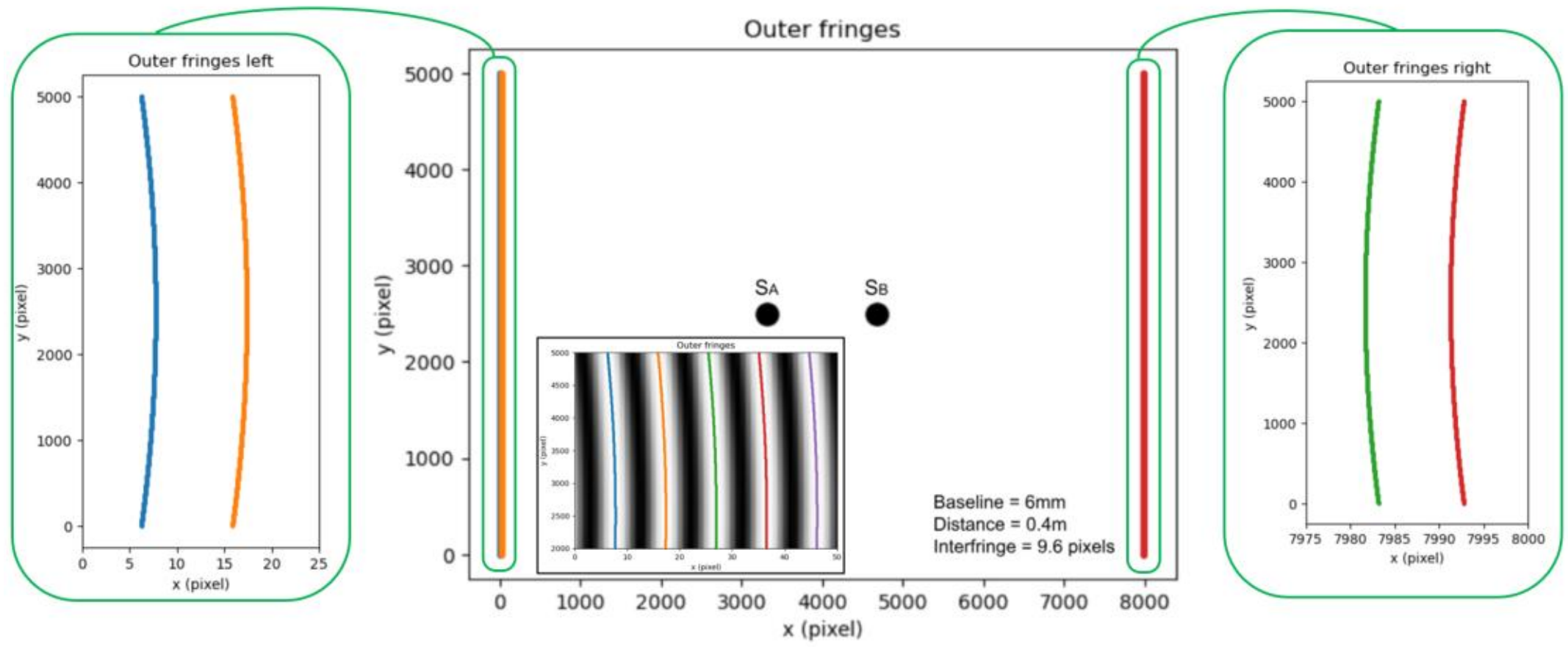

Figure 16. Vertical fringes crests paths detection on a simulated fringe frame. Center: full frame 46 Mpix fringes paths for the 4 extreme fringes on left and right, $s_A, s_B$ the projected fibers positions. Overprinted an extract of the fringes image and the detected paths. Left and right: zooms on the extreme left and right fringes paths showing the hyperbolic shape on this typical case of 46 Mpix detector, fibers at 40 cm.

Another approach for computing the initial sub-iteration involves leveraging the geometric pathways of the fringes to extract the positions of the fibers. This approach would enable the suppression of the numerous deep secondary minima of the inverse problem cost function, with the objective of attaining a unique global minimum. A model of specific features of the image of the fringes can be first extracted, and the unknowns can be estimated from these features. Given the smoothness of the distributions of fiber illumination intensities $I_A(r)$ and $I_B(r)$ ($r$ being the pixels positions), the crests (for $k$ even) and the valleys (for $k$ odd) of the fringes appear at OPD constant paths such that:

$$\frac{2\pi}{\lambda}\delta(\theta,r) = \varphi(\theta,r) = \frac{2\pi}{\lambda}(\|r - s_A\| - \|r - s_B\|) - \phi = k\pi \ (\forall k \in \mathbb{Z}) \quad (6)$$

Where $\varphi$ is fringe phase, $r$ are the pixels of crests and valleys positions, $s_{A,B}$ the fibers tips positions, $\phi$ contains the controlled phase modulator dephasing and is dependent on the $k$ numeration convention, $\theta = \{s_A, s_B, \delta_x, \delta_y\}$ the geometrical parameters to estimate. Denoting by $r_{j,k}$ the position of the $j$-th point along the $k$-th crest (if $k$ even) or valley (if $k$ odd), the parameters can be estimated by a least squares fit:

$$\breve{\theta} = \text{argmin}_\theta \sum_{k\in\mathbb{K}} \sum_{j\in\mathbb{J}_k} \left(\varphi(\theta, r_{j,k}) - k\pi\right)^2 \quad (7)$$

With $\mathbb{K} \subset \mathbb{Z}$ the subset of integers $k$ for which crests and valleys have been detected and $\mathbb{J}_k$ the list of indices $j$ of extracted points along the $k$-th fringe feature (crest or valley). The most informative fringes will be the farther from the fibers baseline projected center. Figure 16 shows the extraction of the fringe paths on a simulated fringe frame for a typical case (GIGAPYX-4600 detector, fibers at 40 cm distance, baseline of 6 mm).

Figure 17 shows this cost function with regard to the distance $D$ (between the center of the baseline and the detector) and baseline $B$ (separation between the fibers) for a simulated fringe frame without noise. Those parameters are expected to be the most correlated ones. The cost function exhibits a global minimum. Therefore, the initial objective has been achieved. The red cross indicates the estimation of the parameters $B$ and $D$ when minimizing the cost function. Therefore, by employing a least-squares algorithm to identify the minimum in Eq. (7) we are able to accurately determine the parameters $B$ and $D$ of the fibers in a simulation devoid of noise. This demonstrates the effectiveness of the proposed method and serves as a foundation for its subsequent implementation on noisy data and actual measurements. Our goal is to replace the first sub-iteration in Fig. 6 by this algorithm which converge quickly and does not show any local minima.

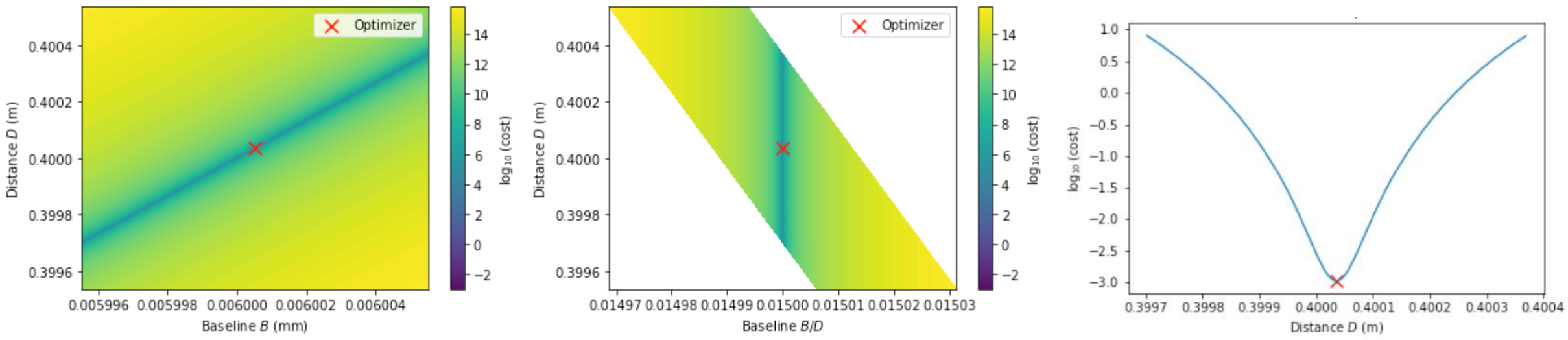

Figure 17. Map of the cost function defined in Eq. (7) in function of the distance of the fibers to the detector and the baseline length corresponding to the separation between the fibers. The two parameters are highly correlated in the left part of the figure, but if we choose as a parameter the baseline normalized by the distance, then B/D is well defined in the center part of the figure. The valley represented in the right part of the figure shows that the value of the best distance is well determined.

## 7. DISCUSSION AND PERSPECTIVES

In the future, the focus should be directed towards enhancing the frame rate of the detector. We can collect $3.5\times10^5$ $e^-/s$ (50ke$^-$ full well @ 7Hz), while the previous studies on 80 × 80 E2V CCD39 could collect $2\times10^8$ $e^-/s$ (200 ke$^-$ full well @ 1 kHz), corresponding to a ratio 560. Therefore, it can be deduced that the $4\times10^{-4}$ pix precision obtained in[9] on analogous hardware would require 2 hours to acquire for one of the two directions. Our Allan deviation study appears to indicate that shorter times (30 min – 1h for $4\times10^{-4}$ pix precision) would be necessary. By approaching the GIGAPYX-4600 frame-rate limit from the datasheet (152 fps), the same precision could be achieved in 6 minutes.

The precision expected from Allan deviation studies are consistent with those established in prior studies[9], yet our simulations indicate an order of magnitude discrepancy. Subsequent research is necessary to ascertain whether this discrepancy originates from the bench stability or from the simulation models. The subsequent step will entail the execution of measurements on the same detector area for two distinct geometrical configurations. This will allow for the assessment of the measurement's repeatability.

In this study, we consider a single illumination wavelength, 632 nm. However, it is important to note that the IPRF (Intra Pixel Response Function) shape is expected to undergo changes in accordance with the illumination wavelength. This is particularly relevant when considering the high precision that is required. Subsequent studies will need to assess this matter with multiple illumination wavelengths, given that the star illumination will be polychromatic.

We conducted both simulations and measurements to assess our capability to map the pixels barycentric response positions on a focal plane array to perform space-based high-accuracy astrometry. It is concluded that, in order to achieve adequate precision of the pixel maps ($5\times10^{-5}$ pix, thus 220 pm), approximately 20,000 frames are required with the detector in its current state to minimize the photon noise contribution to the SNR (signal-to-noise ratio).

The RETINA project[14,15] is currently engaged in the design of the on-board processing capability, which is expected to be highly demanding due to the high number of frames and the nearly 1 Gpix focal plane of the mission.

In the scope of this study, we expect to have reach a $5\times10^{-4}$ pix precision. This precision is predicated on Allan deviation studies and must be substantiated through dedicated measurements to ensure the absence of systematic biases in the estimated $(\delta_x, \delta_y)$ maps.

To address this challenge, a novel estimation algorithm was developed to address the hyperbolicity of the fringes that emerge in the presence of large focal planes. The objective of this study is to validate our capability to reach fine precisions on the 46 Mpix detector and to extend it to the mission 1 Gpix focal plane in the future.

## ACKNOWLEDGEMENTS

The technology maturation program RETINA receives support from the French Government through the France 2030 investment plan, operated by the French National Research Agency under grant ANR-25-EXOR-0003. The work has been also partially supported by the LabEx FOCUS by the French National Research Agency through the grant ANR-11-LABX-0013 and within the framework of the "France 2030" program by the grant ANR-15-IDEX-02. The authors acknowledge financial support from the Centre national d'études spatiales (CNES), France (ROR: https://ror.org/04h1h0y33), thanks to the Research and Technology (R&T) and mission preparation programs. ML acknowledges the support of her PhD grant from CNES and Pyxalis.

This research has made use of data obtained from or tools provided by the portal exoplanet.eu of The Extrasolar Planets Encyclopaedi. This research has made use of the Astrophysics Data System, funded by NASA under Cooperative Agreement 80NSSC21M0056 and the TOPCAT software[17].

## REFERENCES

[1] Malbet, F. and Sozzetti, A., "Astrometry as an Exoplanet Discovery Method," [Handbook of Exoplanets], H. J. Deeg and J. A. Belmonte, Eds., Springer International Publishing, Cham, 689–704 (2018).

[2] Lizzana, M., "High precision astrometry for exoplanets," thesis (2026).

[3] Michelot, J., Douix, M., Mancini, J.-B., Guillon, M., Melendez, K., Ravinet, C., Jouans, M., Estaves, G., Marec, R., Demiguel, S. and Virmontois, C., "GIGAPYX SENSOR PERFORMANCE IN SPACE ENVIRONMENTS."

[4] Lizzana, M., Rousset, H., Soler, S., Pancher, F., Thiébaut, E. and Malbet, F., "Characterization of the GIGAPYX-4600 CMOS visible image sensor in the context of high precision astrometry," JATIS submitted (2026).

[5] Shaklan, S. B., Shao, M., Gursel, Y. and Yu, J., "Measuring Pixel-Position Errors On CCD Imagers," NASA Tech Briefs **19**(10) (1995).

[6] Malbet, F., Léger, A., Shao, M., Goullioud, R., Lagage, P.-O., Brown, A. G. A., Cara, C., Durand, G., Eiroa, C., Feautrier, P., Jakobsson, B., Hinglais, E., Kaltenegger, L., Labadie, L., Lagrange, A.-M., Laskar, J., Liseau, R., Lunine, J., Maldonado, J., et al., "High precision astrometry mission for the detection and characterization of nearby habitable planetary systems with the Nearby Earth Astrometric Telescope (NEAT)," Exp Astron **34**(2), 385–413 (2012).

[7] Shao, M., Zhai, C., Nemati, B., Hahn, I., Trahan, R. and Turyshev, S., "Micro-arcsecond Astrometry Technology: Detector and Field Distortion Calibration," PASP **135**(1049), 074502 (2023).

[8] Nemati, B., Shao, M., Zhai, C., Erlig, H., Goullioud, R. and Wang, X., "Micropixel-level image position sensing testbed," presented at SPIE Optical Engineering + Applications, 8 September 2011, 81510W.

[9] Crouzier, A., Malbet, F., Henault, F., Léger, A., Cara, C., LeDuigou, J. M., Preis, O., Kern, P., Delboulbe, A., Martin, G., Feautrier, P., Stadler, E., Lafrasse, S., Rochat, S., Ketchazo, C., Donati, M., Doumayrou, E., Lagage, P. O., Shao, M., et al., "A detector interferometric calibration experiment for high precision astrometry," A&A **595**, A108 (2016).

[10] Pancher, F., Soler, S., Malbet, F., Rousset, H., Lizzana, M., Léger, A., Lepine, T., Ardellier-Desages, F., Amiaux, J. and Lagage, P.-O., "Setup and performance of focal-plane characterization benches for high-precision astrometry," SPIE **In this volume**(14145–188) (2026).

[11] Moore, A. C., Ninkov, Z. and Forrest, B., "QE Overestimation and Deterministic Crosstalk Resulting from Inter-pixel Capacitance."

[12] Donlon, K., Ninkov, Z. and Baum, S., "Point-spread Function Ramifications and Deconvolution of a Signal Dependent Blur Kernel Due to Interpixel Capacitive Coupling," PASP **130**(989), 074503 (2018).

[13] Le Graët, J., Secroun, A., Barbier, R., Gillard, W., Clémens, J.-C., Conseil, S., Escoffier, S., Ferriol, S., Fourmanoit, N., Kajfasz, E., Kermiche, S., Kubik, B., Smadja, G. and Zoubian, J., "Euclid Near Infrared Spectro-Photometer: spatial considerations on H2RG detectors interpixel capacitance and IPC corrected conversion gain from on-ground

characterization," X-Ray, Optical, and Infrared Detectors for Astronomy X, A. D. Holland and J. Beletic, Eds., 98, SPIE, Montréal, Canada (2022).

[14] Amiaux, J., Malbet, F., Ardellier-Desages, F., Bendek, E. A., Doumayrou, E., Frugier, P.-A., Goullioud, R., Greene, T., Lagage, P. O., Lizzana, M., Martignac, J., Michelot, J., Pancher, F., Pichon, T., Roberge, A., Ronayette, S., Rousset, H. and Sitarski, B. N., "High-precision high-accuracy astrometry for the Habitable World Observatory," SPIE **In this volume**(14145–67) (2026).

[15] Ardellier-Desages, F., Amiaux, J., Doumayrou, E., Frugier, P.-A., Lagage, P. O., Martignac, J., Pichon, T., Ronayette, S., Malbet, F., Pancher, F., Lizzana, M., Rousset, H. and Michelot, J., "Technological maturation of a CMOS gigapixels astrometry instrument for Habitable World Observatory space mission," SPIE **In this volume**(14145–66) (2026).